\documentclass[conference]{IEEEtran}
\ifCLASSINFOpdf
  \usepackage[pdftex]{graphicx}
\else
  \usepackage[dvips]{graphicx}
\fi
\begin{document}
\onecolumn
\vspace{20mm}
{\huge \bf Similarities and differences between stimulus tuning in the inferotemporal visual cortex and convolutional networks.}
\vspace{10mm}

{\large Bryan Tripp, University of Waterloo, Canada}
\vspace{10mm}

{\large Submitted on December 1, 2016 to IJCNN 2017.}
\vspace{10mm}

{\large © 2016 IEEE. Personal use of this material is permitted. Permission from IEEE must be obtained for all other uses, in any current or future media, including reprinting/republishing this material for advertising or promotional purposes, creating new collective works, for resale or redistribution to servers or lists, or reuse of any copyrighted component of this work in other works.}

\clearpage
\twocolumn
%
\title{Similarities and differences between stimulus tuning in the inferotemporal visual cortex and convolutional networks.}

\author{\IEEEauthorblockN{Bryan P. Tripp}
\IEEEauthorblockA{Department of Systems Design Engineering \&
Centre for Theoretical Neuroscience\\
Waterloo, Ontario, Canada\\
Email: bptripp@uwaterloo.ca}}


%


\maketitle

\begin{abstract}
Deep convolutional neural networks (CNNs) trained for object classification have a number of striking similarities with the primate ventral visual stream. In particular, activity in early, intermediate, and late layers is closely related to activity in V1, V4, and the inferotemporal cortex (IT). This study further compares activity in late layers of object-classification CNNs to activity patterns reported in the IT electrophysiology literature. There are a number of close similarities, including the distributions of population response sparseness across stimuli, and the distribution of size tuning bandwidth. Statisics of scale invariance, responses to clutter and occlusion, and orientation tuning are less similar. Statistics of object selectivity are quite different. These results agree with recent studies that highlight strong parallels between object-categorization CNNs and the ventral stream, and also highlight differences that could perhaps be reduced in future CNNs.
\end{abstract}


\section{Introduction}
Deep convolutional neural networks (CNNs) that are trained for object categorization have activity that resembles that of the ventral visual stream in several respects. For example, early layers often exhibit Gabor-like receptive fields similar to V1 \cite{Zeiler2014}, and separation of early layers across multiple GPUs has resulted in separation of color and edge encoding \cite{Krizhevsky2012}, analogous to V1 blobs and interstripes. 

Furthermore, internal activity in later layers of such CNNs is closely related to activity later in the primate ventral visual stream. \cite{Yamins2014} showed that much of the variance in activity of IT neurons (up to about 30\%) could be accounted for by multilinear regression with the responses of units in the CNN's second-last layers. They also showed that these linear predictions had IT-like correlations related to object categories \cite{Yamins2014}. Models that have been specifically developed to approximate the ventral stream, including HMAX \cite{Serre2007} and VisNet \cite{Rolls2012}, have lacked such realistic correlations \cite{Khaligh-Razavi2014}. \cite{Yamins2014} also showed that activity in intermediate layers of these CNNs predicted activity of neurons in V4, which is a major source of input to IT \cite{Markov2014}. 

Many IT neurons are relatively invariant to scale and translation changes \cite{Schwartz1983}, and a few are also invariant to orientation changes. Consistent with this, \cite{Zeiler2014} found that the vector of population responses in the second-last later of a CNN changed relatively little with with position, size, and orientation, compared to responses in the first layer.  

More recently, \cite{Hong2016} found that IT neurons and units of object-classification CNNs provided comparable information about category-orthogonal object properties, including object position and orientation. 

Importantly, CNNs perform similarly to humans \cite{Krizhevsky2012,Zeiler2014} in ``core object recognition'', a rapid feedforward process that allows primates to recognize objects in natural scenes after seeing them very briefly \cite{dicarlo2012does}, on similar timescales to inter-saccade intervals. 

Core object recognition is a sophisticated function, and a major function of the ventral visual stream. It is highly relevant to neuroscience that this this function can now be performed by artificial systems. CNNs can not be considered realistic models of the ventral stream, because they learn in very different ways, consist of much simpler units, typically lack lateral connections, etc. \cite{Robinson2015}. However, the fact that they have realistic core object recognition capabilities, and also exhibit internal representations that resemble those in the ventral stream, constrains ideas about the roles of these missing details in vision. 

\subsection{Purpose of this study}

Linear combinations of CNN activities are currently the best predictors of IT activity in high-variation stimulus conditions, but they leave about half the variance of IT responses unaccounted for. It would be interesting to know whether CNNs predict certain factors of variation in IT better than others. Also, the CNN activities themselves (as opposed to optimal linear combinations of them) have somewhat different correlation patterns \cite{Yamins2014,Khaligh-Razavi2014}.

The purpose of the present study is to compare the response properties of deep object-recognition CNNs and IT neurons, along several dimensions of stimulus variation that have been explored in IT. 
Remarkably, despite the optimization of both IT and CNNs for natural scenes, both systems discriminate the same simplified stimuli (e.g. line drawings) in a similar manner (see Results). This study focuses on tuning curves of individual units, allowing more direct comparison with IT than previous investigations of population response vectors \cite{Zeiler2014}. 

To summarize the results, many response properties of the CNNs are strikingly similar to previously reported responses of IT neurons. These including the distributions of population sparseness across stimuli, and the distribution of size-tuning bandwidth. Other properties, including responses to partial occlusion and clutter, orientation tuning curves, and patterns of size invariance, were somewhat different. Distributions of object selectivities over neurons were quite different. Along with differences in correlations of IT responses within and between object categories \cite{Yamins2014,Khaligh-Razavi2014}, these results help to clarify the relationships between stimulus representation in object-classification CNNs and IT. 

Preliminary versions of some of these analyses appear in \cite{Khan2017}. 

\section{Methods}
Some of the figures replot responses of IT neurons from the experimental literature. The data points for these plots were extracted from published figures using Web Plot Digitizer (http://arohatgi.info/WebPlotDigitizer/).

\subsection{Networks}
We study responses of two well-known deep convolutional networks. Both networks were trained to classify objects in natural scenes for the ImageNet Large-Scale Visual Recognition Challenge \cite{Russakovsky2015}. One was Alexnet \cite{Krizhevsky2012}. This network has previously been compared with IT in \cite{Cadieu2014} and \cite{Khaligh-Razavi2014}. 
The other network, VGG-16 \cite{simonyan2014very}, has more layers and smaller kernels. Both of the implementations used here were taken from https://github.com/heuritech/convnets-keras. The Alexnet implementation was adapted from a version trained for \cite{Ding2015}. The VGG-16 implementation was adapted from a version from https://gist.github.com/baraldilorenzo, which was in turn adapted from the version provided by \cite{simonyan2014very}. 

Previous comparisons of CNNs with IT \cite{Yamins2014,Khaligh-Razavi2014,Hong2016} have focused on the second-last network layer, a fully-connected layer that precedes the final softmax classification layer. This layer is in a similar position to the inferotemporal cortex (IT) in the ventral stream, the neurons of which are important for visual object categorization, but also contain much non-category information, e.g. \cite{Schwartz1983,Hong2016}. For comprehensiveness, we also consider responses from the surrounding layers, i.e. the last and third-last layers, which turn out to more closely resemble IT in some respects.

\subsection{Stimulus Images}
Stimulus images were obtained from several sources. To investigate stimulus selectivity and population response sparseness, we used a set of 806 images developed by \cite{Lehky2011} and used for the same purpose in IT. The images were extracted directly from their supplementary materials file. The extracted images had a lower resolution (60 by 60 pixels) than the authors reported using (125 by 125 pixels). However, the objects were easily identifiable. The extracted images were embedded within a larger gray background. The extracted images themselves had light borders and somewhat noisy gray backgrounds. The borders were removed by omitting the outer two pixels, and variations were reduced in the backgrounds by setting to a uniform gray value ([red,green,blue]=$[128,128,128]^T$) any vector with a cartesian distance of less than 15 from this value. 

A few additional higher-resolution images (banana, car, shoe, etc.) were downloaded from various internet sources and given uniform gray backgrounds. These images were used to examine orientation, position, and size tuning. Images for studying responses to rotation-in-depth were rendered in SketchUp. 3D scooter and human head models were obtained from the SketchUp 3D Warehouse. A wire-like object model, as in \cite{Logothetis1995} was built in OpenSCAD. Other images were generated using custom code in order to approximate stimuli in \cite{Kovacs1995}. Finally, some stimulus images were copied from figures in \cite{Schwartz1983} and \cite{Tanaka2003}. 


\section{Results}
\subsection{Stimulus selectivity and population sparseness}
Two fundamental statistical properties of IT population responses are 1) the sparseness of responses across the population to various stimuli, and 2) the sparseness of individual neurons' responses evaluated over multiple stimuli (sometimes called stimulus selectivity). Both have been described in terms of excess kurtosis (kurtosis in excess of that of the Gaussian distribution).  
\cite{Lehky2011} compared stimulus selectivity and population sparseness in a large dataset that included responses of 674 IT neurons to 806 different images. They found that distributions of selectivity and sparseness had a similar shape, and that sparseness exceeded selectivity. They found that this difference could be reproduced in a simple statistical model, as a result of non-uniform mean spike rates across the population.  

\begin{figure}
\begin{centering}
\includegraphics[scale=0.6]{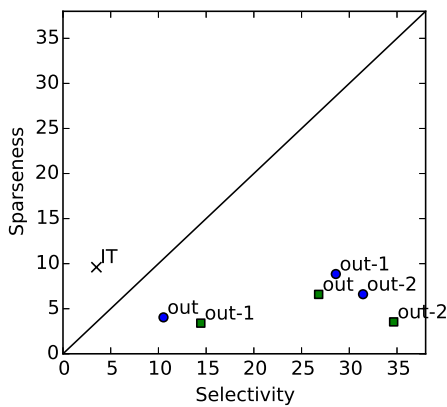}
\par\end{centering}
\caption{Sparseness vs. selectivity of IT neurons (from \cite{Lehky2011}), and of CNN units responding to a similar group of stimuli. Sparseness exceeds selectivity in IT \cite{Lehky2011}, whereas selectivity consistently exceeds sparseness, in each layer of both CNNs. The blue circles are Alexnet layers and the green squares are VGG-16 layers. The labels indicate the distance from the output, i.e. ``out'' is the last layer, ``out-1'' is the second-last, and ``out-2'' is the third-last. }
\label{fig:cnn-selectivity}
\end{figure}

Datasets similar to that of \cite{Lehky2011} were collected from the CNNs, using approximately the same set of images (see Methods). Figure \ref{fig:cnn-selectivity} plots selectivity vs. sparseness of units in the final three layers of both CNNs (Alexnet and VGG-16). 674 units were chosen from each layer, to match the size of the dataset in \cite{Lehky2011}. The 674 units with the highest peak activations in response to the stimulus images were chosen from each layer. The sparseness of the CNN activity was similar to that of the IT neurons. However, in contrast to IT neurons, each of the CNN layers exhibited higher selectivity than sparseness. The sparseness of the CNN activity was disproportionately due to very rare activation of some units, while the remaining units were activated less sparsely than average. The sparseness distributions had a similar shape to that of \cite{Lehky2011}, but the selectivity distributions had longer tails (not shown).

\subsection{Orientation Tuning}
The responses of most IT neurons varies strongly with orientation in each dimension \cite{Logothetis1995}, although a small fraction ($<1^{\circ}$ in \cite{Logothetis1995}) are essentially orientation-invariant. Responses are typically strongest around a single preferred orientation unless the object is rotationally symmetric. Orientation tuning curves often resemble gaussian functions. \cite{Logothetis1995} reported somewhat narrower tuning curves for rotation in depth than rotation in the image plane. For neurons that responded to faces, \cite{hasselmo1989object} reported a fairly uniform distribution of preferred viewing angles between left and right profiles (their Figure 11). 

\begin{figure*}
\begin{centering}
\includegraphics[scale=0.45]{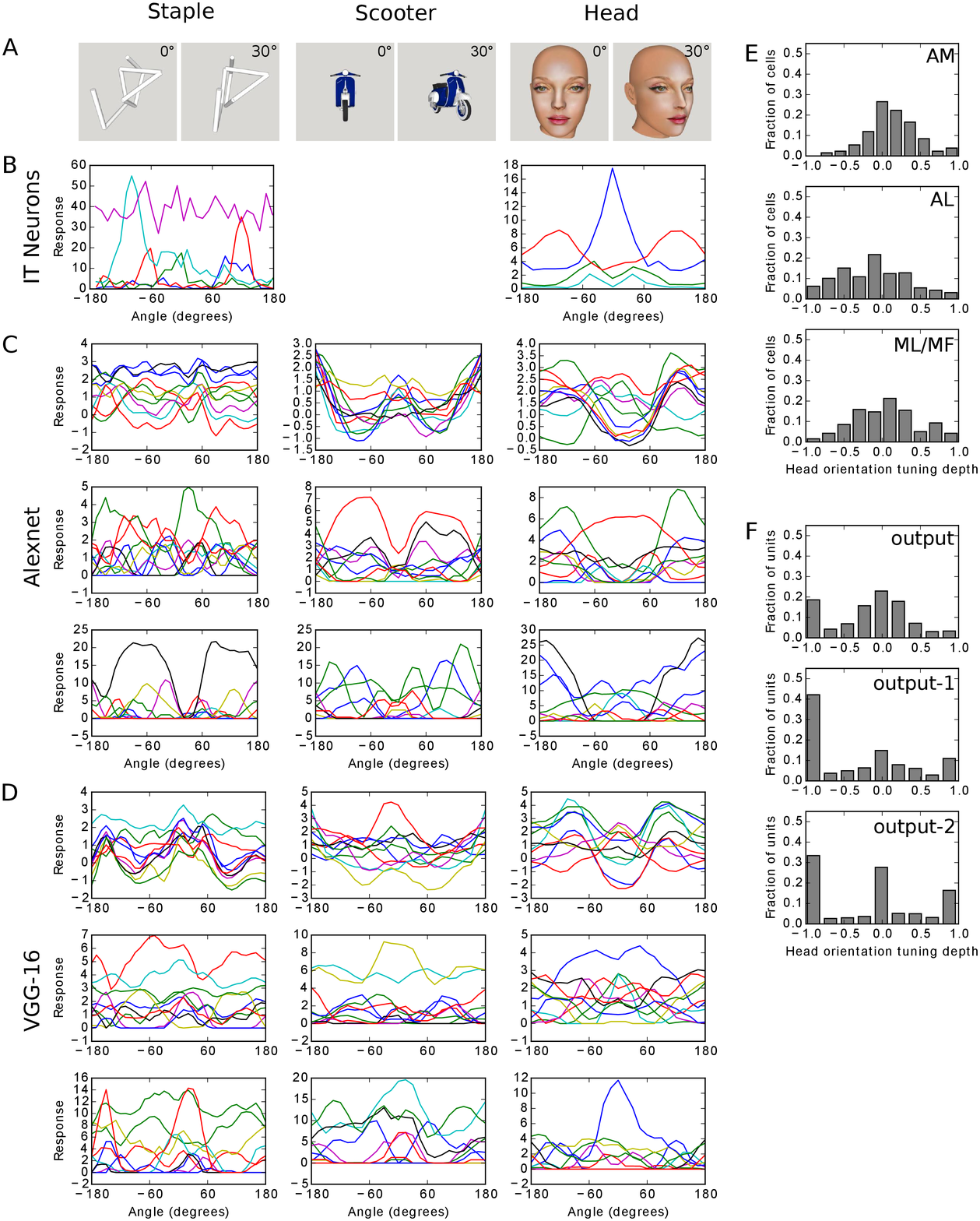}
\par\end{centering}
\caption{Tuning curves for rotation-in-depth of a wire-like object (first column), a scooter (second column), and a human head (third column). A) Example views that were used as inputs to the CNNs. B) Tuning curves of IT neurons in response to similar (not identical) objects. Responses to a wire-left object (left) and head (right) are redrawn from data in  \cite{Logothetis1995} and \cite{Freiwald2010}, respectively. C) Responses of units in the final three layers of Alexnet. The columns correspond to wire, scooter, and head stimuli. The top row shows responses from the output layer, and the other rows show responses from the immediately preceeding layers. In each case the responses of the first ten units with peak responses greater than two are shown, and the tuning curves are smoothed by averaging over groups of three neighbouring points. D) Same format as C for VGG-16. E) Histograms of tuning depth of IT neurons in different face-responsive areas, replotted from \cite{Freiwald2010}. The tuning depth is the frontal response minus the profile response, divided by their sum. Depths are closer to zero (tuning curves are flatter) in the anterior medial area (AM) relative to more posterior areas. F) Depth histograms of head tuning curves from the final three layers of Alexnet. Analogous to the neural data, depth is lowest in the output layer. In contrast with the neural data, the distributions in other layers are multimodal.}
\label{fig:cnn-orientation3d}
\end{figure*}

Figure \ref{fig:cnn-orientation3d} shows tuning curves over rotations in depth about the vertical axis. Figure \ref{fig:cnn-orientation3d}A shows example views of three stimulus objects: a ``wire-like'' object, similar to those used in \cite{Logothetis1995}; a more complex inanimate object that might be encountered in life (a scooter); and a human head.
Figure \ref{fig:cnn-orientation3d}B shows examples of IT-neuron orientation-tuning curves for a wire-like object (replotted from \cite{Logothetis1995}) and a human head (replotted from \cite{Freiwald2010}). The wire (left) elicited narrow, roughly Gaussian tuning curves from most neurons \cite{Logothetis1995}. A minority of neurons responded strongly to roughly mirror-symmetric views separated by 180$\deg$. Finally, a very small minority of neurons had approximately view-invariant responses. 

The left columns of Figure \ref{fig:cnn-orientation3d}C and D show responses of units in Alexnet and VGG-16, respectively. In contrast with IT neurons, the CNN units typically have multimodal tuning curves. Units in the last two layers have broad tuning. However, the third-last layer of Alexnet has several fairly narrow, roughly Gaussian tuning curves that resemble view-selective neurons in \cite{Logothetis1995}, except that they are bimodal (perhaps due to use of flipped images to enhance the dataset in training).

As the wire-like object is unlike objects that the networks encountered in training, orientation tuning curves were also plotted for an object from one of the ImageNet categories (a scooter). As with the wire, most responses in both Alexnet and VGG-16 were broad and multimodal. There are no examples of narrow uni- or bimodal Gaussian tuning in this column.

The right column of \ref{fig:cnn-orientation3d}B shows responses of several cells in face-selective regions of IT from \cite{Freiwald2010}. Some of these cells preferred front views, and some preferred profile views, but responded similarly to left and right profiles. Interestingly, CNN tuning curves also showed these two patterns, although with somewhat broader tuning. 

\begin{figure}
\begin{centering}
\includegraphics[scale=0.5]{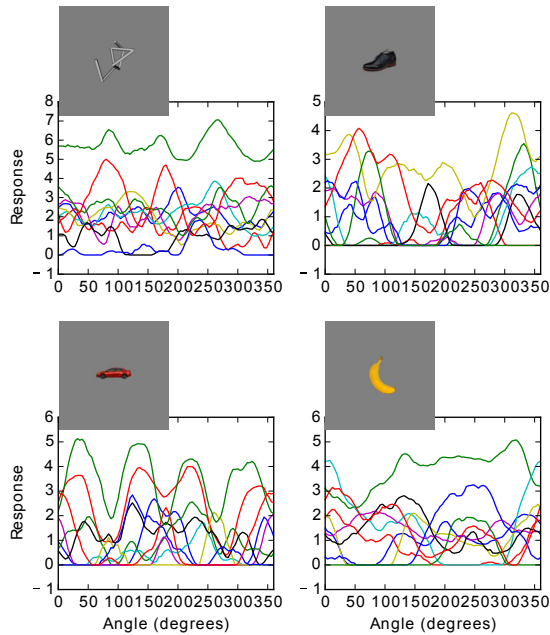}
\par\end{centering}
\caption{Tuning curves for image-plane orientation of staple, shoe, car, and banana images. In each case, the responses of the first ten units with peak responses $>2$ are shown. The plotted tuning curves were smoothed by averaging over groups of four neighbouring points, so that the different curves are easier to distinguish, according to the Gestalt law of good continuation.}
\label{fig:cnn-orientation}
\end{figure}

In \cite{Logothetis1995}, rotations in the image plane produced responses that were similar to rotation in depth (many neurons had roughly Gaussian tuning) except that tuning was somewhat broader. Figure \ref{fig:cnn-orientation} shows examples of image-plane orientation tuning curves from the second-last layer of Alexnet. In contrast with typical IT responses to image-plane rotation, most of the CNN responses are multimodal. 

\subsection{Size Tuning and Invariance}

\begin{figure}
\begin{centering}
\includegraphics[scale=0.36]{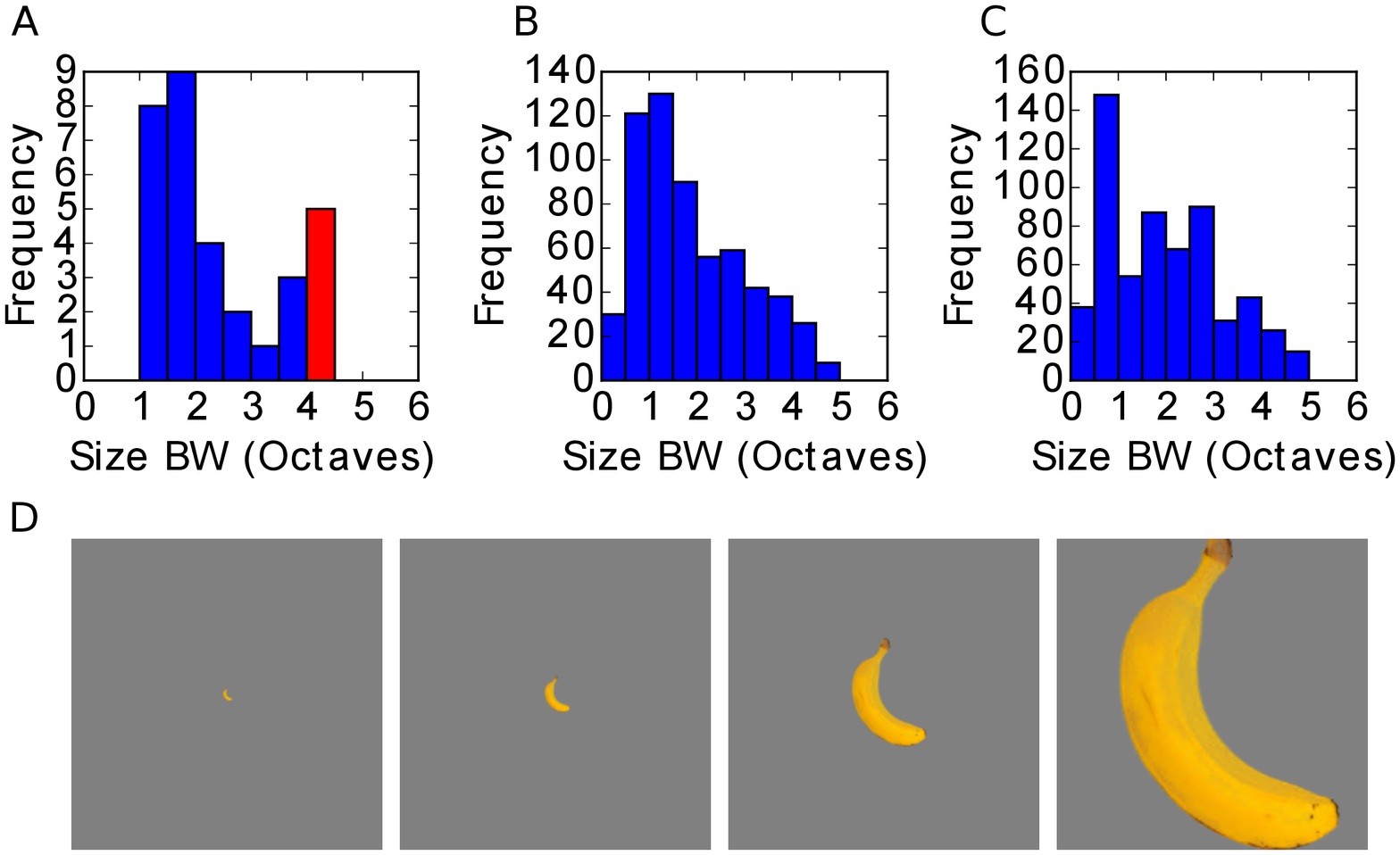}
\par\end{centering}
\caption{A) Histogram of half-magnitude size tuning bandwidths, replotted from \cite{Ito1995}. The range of stimulus scales did not fully span the half-magnitude ranges of the units plotted in red. B) Histogram of half-magnitude bandwidths for Alexnet. Stimuli included scaled versions of banana, shoe, and car images. For each object, the 200 units with the strongest peak responses were included in the histogram. C) As B, but for VGG-16. D) Examples of banana images at some of the different scales used with the CNNs.}
\label{fig:cnn-size}
\end{figure}

IT neurons are also tuned for size. \cite{Ito1995} studied the distribution of size tuning bandwidths, i.e. the ratio between the larger and smaller sizes at which the response drops to half it's peak. The mode of this distribution was approximately 1.5 octaves (their Figure 8), and bandwidths of up to 4 octaves were reported. 
Figure \ref{fig:cnn-size}B and C shows histograms of size tuning bandwidths in Alexnet and VGG-16, respectively. They are both quite similar to the distribution in IT. For example the Alexnet histogram has a mode around 1 (versus 1.5 in IT), and values up to 4.6 (vs. a few values at or above 4, the highest value measured, in IT).  

\begin{figure}
\begin{centering}
\includegraphics[scale=0.43]{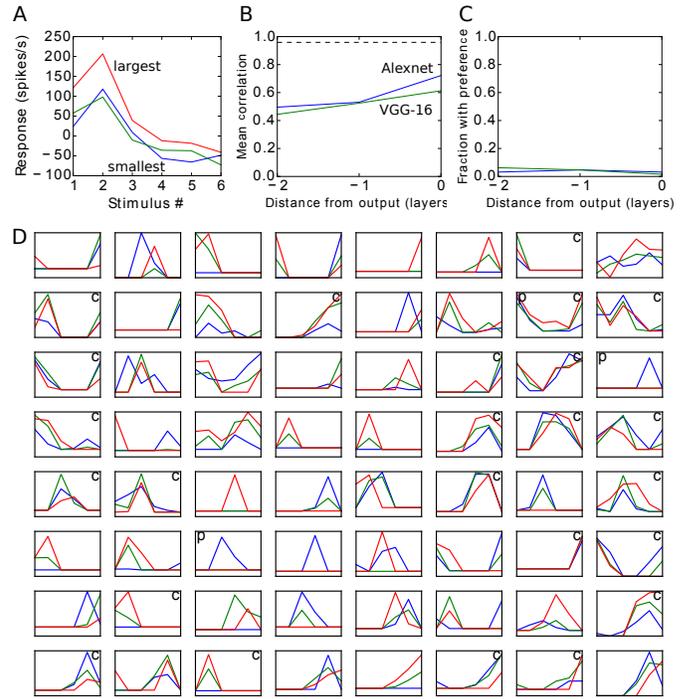}
\par\end{centering}
\caption{Size invariance in the CNNs, examined with stimuli from \cite{Schwartz1983}. A) Responses of an example IT neuron, replotted from \cite{Schwartz1983}. The horizontal axis spans different stimulus shapes, the vertical axis is the mean spike rate. The red, green, and blue curves correspond to small, medium, and large versions of the stimuli. B) The dashed line is the correlation between the tuning curves in A. The blue and green are mean correlations of similar tuning curves in Alexnet and VGG-16. The correlations are plotted versus layer, with the output layer on the right. C) The fraction of units in each CNN layer that had the same scale preference for all shapes (like the example neuron in A, in which the red curve does not cross the others). D) Examples of tuning curves in the second-last layer of Alexnet, with the same format as A. Many of the units' responses are correlated across scales (those with mean correlations >0.75 are marked `c'). However, few have a uniform scale preference across shapes (marked 'p'). The units in B-D were the first 64 units in each layer with peak responses $>2$ over all stimuli. 
}
\label{fig:cnn-size-invariance}
\end{figure}

Responses of the majority of IT neurons exhibit size invariance \cite{Schwartz1983}. Spike rates vary with stimulus size, but the relative spike rates in response to different stimuli are similar regardless of sizes. A prototypical example is shown in Figure \ref{fig:cnn-size-invariance}A. The neuron's tuning curve over six different shapes is roughly maintained over different stimulus scales. However, the gain of this curve depends on the scale. Figure \ref{fig:cnn-size-invariance}D shows responses of units in the second-last layer of Alexnet. The stimulus shapes and relative scales are the same as in \cite{Schwartz1983}. Similar to the IT neurons, many of the CNN units' tuning curves have similar shapes when the stimuli are shown at different scales. A `c' in the upper-right of each plot indicates that the mean correlation between curves is greater than 0.75. However, only a small minority of CNN units have a clear preferred size (marked with `p' in the upper left) like the canonical example of \cite{Schwartz1983}.

Panels \ref{fig:cnn-size-invariance}B and C show that these results are similar for other layers of Alexnet and VGG-16. Mean correlations are well below those of \ref{fig:cnn-size-invariance}A, and only a small fraction of CNN units in any layer show a clear preference for a certain size that is maintained across multiple shapes. Shape and size tuning may be somewhat less separable in the CNNs than in IT. 

\subsection{Position Tuning and Invariance}
\begin{figure}
\begin{centering}
\includegraphics[scale=0.4]{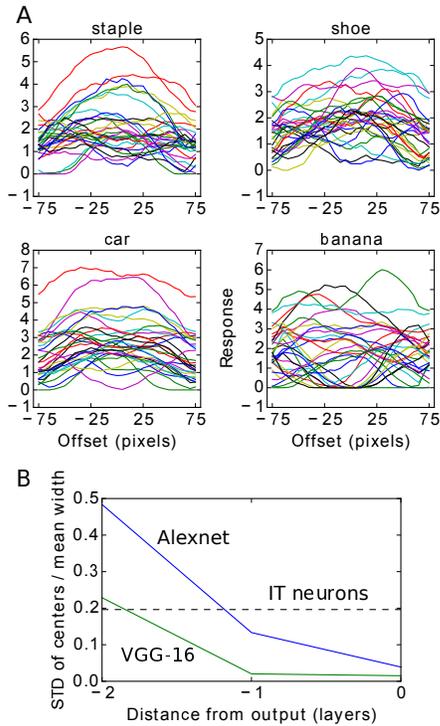}
\par\end{centering}
\caption{A) Responses to staple, shoe, car, and banana images translated to different horizontal positions within a gray background. The offset is the distance from centre, in an image that was 256 pixels wide. At the largest offsets, the stimuli nearly reached the edge of the background. B) Ratios of standard deviation of centres of mass, over mean half-width, of position-tuning curves. The dashed line is the ratio from data in \cite{OpDeBeeck2000}. The blue and green lines are calculated from the responses of Alexnet and VGG-16 to the translated staple, shoe, car, and banana images.}
\label{fig:cnn-position-tuning}
\end{figure}

Many IT neurons exhibit translation invariance as well as size invariance \cite{Ito1995}. 
Figure \ref{fig:cnn-position-tuning}A shows examples of response variations with horizontal stimulus position, in the second-last layer of Alexnet. In each plot, the first 30 units with responses $>2$ are shown. As in \cite{OpDeBeeck2000}, many of the tuning curves are roughly gaussian, and most units prefer centrally located stimuli. 

The dispersion of preferences around zero (the fovea) cannot be directly compared, because the mapping between pixels and degrees visual angle is not well defined (it is not uniform in the ImageNet training data). However, it is possible to compare the ratio of this dispersion to the widths of the tuning curves. Figure \ref{fig:cnn-position-tuning}B plots ratios of standard deviation of tuning curve centres of mass, divided by mean tuning curve width (distance between points on each side of the centre of mass that fall below half the peak). The dashed line is this ratio taken from the IT data in \cite{OpDeBeeck2000}. The blue and green lines are from Alexnet and VGG-16, respectively. As a fraction of tuning curve width, the position-tuning centres in the last layers of both networks are much more narrowly dispersed than IT position-tuning curves. However, the second-last layer of Alexnet and third-last layer of VGG-16 are similar to the IT neurons in this respect.

\begin{figure}
\begin{centering}
\includegraphics[scale=0.43]{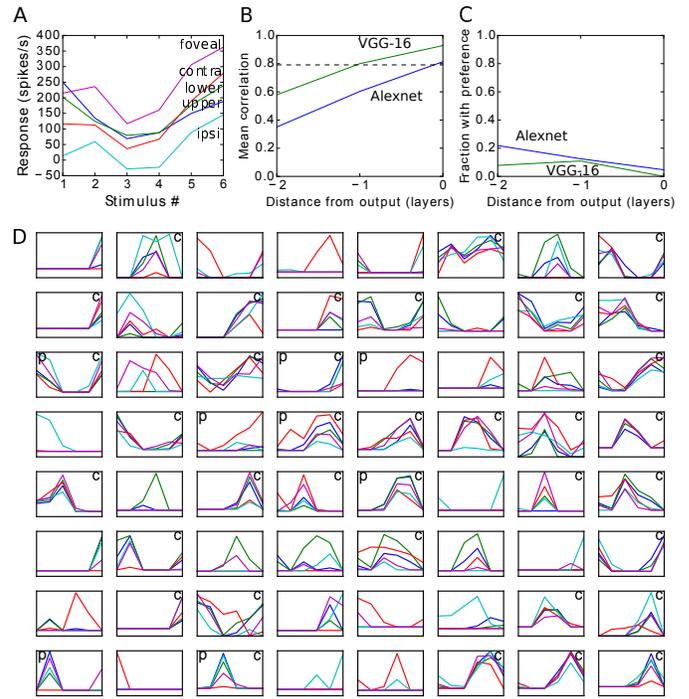}
\par\end{centering}
\caption{Translation invariance in CNN units. Both the plot format and CNN unit selection procedure are the same as in Figure \ref{fig:cnn-size-invariance}. A) Data replotted from an example neuron in \cite{Schwartz1983}. B) Mean correlations of tuning curves in Alexnet (blue) and VGG-16 (green) compared with that of the curves in panel A (dashed). C) Fraction of units in each layer in which the same position was preferred for all but at most one shape, like the IT neuron shown in panel A. D) Example tuning curves from the second-last layer of Alexnet, in the same format as A. The different tuning curves correspond to stimuli at the centre, and translated up, down, left, and right from centre. The translation distance relative to stimulus size was roughly matched to that in \cite{Schwartz1983}.}
\label{fig:cnn-position-invariance}
\end{figure}

Figure \ref{fig:cnn-position-invariance} shows that (in contrast with size invariance) the CNNs have similar position invariance to IT. The last and second-last CNN layers have average correlations in these tuning curves that are similar to the example given by \cite{Schwartz1983} (their Figure 2D), which is consistent with the population averages of \cite{OpDeBeeck2000} (their Figure 10C). A substantial fraction of the units have a position preference that is consistent across shapes, with at most one shape at which the preferred position is inconsistent (as in the IT data in \ref{fig:cnn-position-invariance}A).   

\subsection{Occlusion}
\cite{Kovacs1995} studied variations in the responses of IT neurons as shape stimuli were partially occluded. Neurons maintained their shape preferences, but responded less vigorously as the shapes were occluded more fully. 

\begin{figure}
\begin{centering}
\includegraphics[scale=0.35]{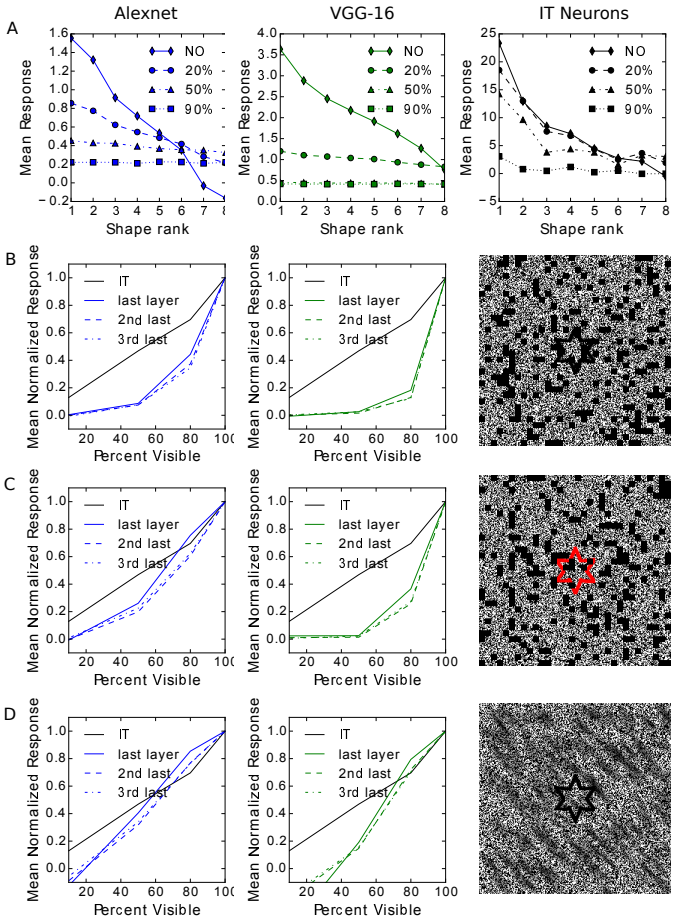}
\par\end{centering}
\caption{Results of an occlusion experiment. A) Shape tuning curves averaged over the 500 units with the strongest responses from the second-last layers of Alexnet (left) and VGG-16 (center), and data from IT neurons in \cite{Kovacs1995} (right). The CNN responses are averages over ten random occlusion patterns (i.e. different random placements of occluding blocks). An example stimulus is shown on the right of panel B. The background, shapes, and occluders were intended to approximate those in \cite{Kovacs1995}. The lines correspond to no occulusion, and occlusion of 20\%, 50\%, and 90\% of the image. These responses have baseline activity subtracted. In \cite{Kovacs1995}, baseline activity was taken from an inter-stimulus delay period. In the CNNs, baseline activity is the response to a black image. B) Responses to preferred stimuli at different occlusion levels. The black, blue, and green lines are IT data from \cite{Kovacs1995}, and unit responses from Alexnet and VGG-16, respectively. These responses were normalized by subtracting the mean response to the least-preferred stimuli, and dividing by the response to the unoccluded preferred stimuli. An example star-shaped stimulus with 20\% occlusion is shown on the right. C) As B, but with stimulus shapes in red rather than black. D) As B, but with the occlusion patterns blurred to approximate motion of the occluder in \cite{Kovacs1995}. }
\label{fig:cnn-occlusion}
\end{figure}

Figure \ref{fig:cnn-occlusion} shows results of a similar experiment with the CNN. For this experiment, following \cite{Kovacs1995}, several shape outlines were drawn on top of a noisy background, and these shapes were occluded by replacing blocks of pixels with black. As with IT neurons, shape preference order tended to be maintained with increasing occlusion, and the amplitude of the preferences was reduced with high occlusion. \ref{fig:cnn-occlusion}A shows tuning curves over different shapes, with different occlusion levels, for the second-last layer of Alexnet (left), the second-last layer of VGG-16 (centre), and IT neurons from \cite{Kovacs1995}. The stimuli used to generate these plots were black outlines, and the occluders were black squares (see example in Figure \ref{fig:cnn-occlusion}B). In these examples, partial occlusion attenuated the CNN unit responses more strongly than the IT responses. For example, at 50\% occlusion, VGG-16 units responded uniformly and minimally regardless of the stimulus shape. 

Figure \ref{fig:cnn-occlusion}B plots mean normalized responses to units' preferred shapes, as a function of the unoccluded percentage of the image. Compared to the IT neurons (black line), responses in each layer of Alexnet (left) and VGG-16 (right) are attenuated much more by occlusion. 

The CNN and IT responses cannot be compared directly, due to temporal evolution of the IT stimulus over time that cannot be reproduced in the CNNs. For example, in the IT experiment, the shape appeared after the occluder, providing an additional segmentation cue. To provide a comparable cue for the CNNs, the simulations were repeated with the shapes and occluders in different colors (right panel of \ref{fig:cnn-occlusion}C). This reduced the attenuation of CNN responses with occlusion. Another complication is that the IT data shown in \ref{fig:cnn-occlusion}A (right) is from experiments in which the occlusion patterns moved. To approximate such motion, the CNN experiment was repeated with occluders that were blurred over a similar distance relative to the stimulus sizes (right panel of \ref{fig:cnn-occlusion}D). This stimulus change also reduced the attenuation of CNN responses with occlusion. In this condition, Alexnet's occlusion responses were similar to those of IT neurons. Notably however, the IT neurons responded similarly to 50\% occlusion regardless of whether it was moving or stationary \cite{Kovacs1995}.  

Relatedly, occlusion of certain parts of objects impairs the performance of CNNs more than occlusion of other parts \cite{Zeiler2014}, consistent with responses in IT \cite{nielsen2006dissociation}.  

\subsection{Clutter}
\begin{figure}
\begin{centering}
\includegraphics[scale=0.35]{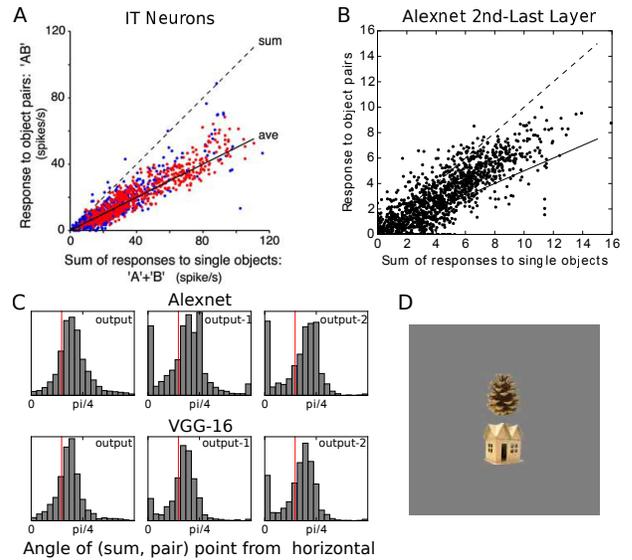}
\par\end{centering}
\caption{Effects of low levels of clutter. A) Responses of IT neurons to pairs of objects (vertical axis) versus sums of their responses to each object presented alone (reproduced with permission from \cite{Zoccolan2005}). Red dots indicate that each stimulus elicited a response greater than the background rate, and blue dots indicate that only one stimulus did so. The responses to pairs of objects are mostly close to the average of responses to each object alone. B) Responses of units in the second-last layer of Alexnet to pairs of objects versus sums of responses to each object alone. Compared to IT, the CNN responses to paired objects are more varied, and are typically greater than the mean of the single-object responses. The plot includes responses of the 100 units with the greatest peak responses across pair conditions. A random selection of units produced a similar pattern but with a higher density of points in the lower left corner. C) Angles of points (counter-clockwise from rightward) in the same space as the scatterplots in panels A an B. Each histogram corresponds to a layer of one of the CNNs. The red lines indicate the angle $\arctan(1/2)$. This corresponds to points on the solid lines in A and B, i.e. responses to paired objects that equal the average of responses to individual objects. The average angle for the IT responses in A is approximately $\arctan(1/2)$. D) An example of a pair of images.}
\label{fig:cnn-clutter}
\end{figure}

\cite{Zoccolan2005} compared the responses of IT neurons to isolated stimulus objects with responses to multiple objects. Responses to pairs of objects were closely clustered around the average response (rather than the sum) of responses to each object individually \ref{fig:cnn-clutter}A.  

A similar experiment was performed with the CNN, using eights images from the dataset of \cite{Lehky2011}. Images were shown at two locations (above and below centre) in a gray background, as in the example of Figure \ref{fig:cnn-clutter}D. Four images were used in the upper location, and four different images were used in the lower location. Unit responses were recorded with each of the upper and lower stimuli alone, and also with each of the 16 possible pairs of upper and lower stimuli. As shown in \ref{fig:cnn-clutter}B, responses of the second-last layer of Alexnet to pairs of images did not cluster tightly around the average of responses to corresponding isolated images (in contrast with \cite{Zoccolan2005}), but were more broadly distributed, and typically intermediate to the sum and mean of the isolated responses. 

Figure \ref{fig:cnn-clutter}B shows histograms of the angles of these points, counter-clockwise from the positive horizontal axis. An angle of $\pi/4$ means that the unit's response to paired images equals the sum of responses to each image alone. An angle of $\arctan(1/2)$ (red lines) means that the unit's response to paired images equals the average of responses to each image alone. In each layer of both networks, the mean and mode of the angles is between $\arctan(1/2)$ and $\pi/4$, suggesting that the CNN responses are not normalized as strongly as IT responses. This is true even in the output layers, which have softmax nonlinearities. 

\subsection{Strong Responses to Simplified Images}

Tanaka and colleagues found that many IT neurons that responded strongly to complex images also responded strongly to highly simplified versions of these images \cite{Tanaka2003}. After showing a monkey many objects in a standard set, and determining the most effective stimulus from the set (i.e. the one that elicited the strongest response), the most effective stimulus was progressively simplified to determine which features were necessary for maximal activation. For example, a cell that responded maximally to a sketch of a striped cat also responded maximally to a sketch of a striped ball on top of another striped ball. Response strength often increased with simplification. IT neurons typically responded maximally to moderately complex stimuli that were more complex than primitive shapes, but less complex than natural objects. 

\begin{figure}
\begin{centering}
\includegraphics[scale=0.5]{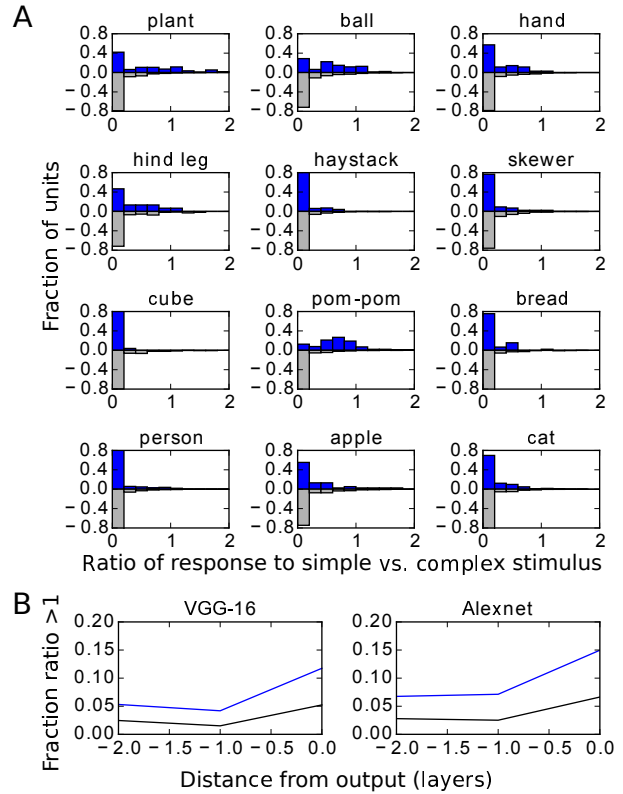}
\end{centering}
\caption{Responses of CNN units to simplified versions of most effective stimuli. A) Responses of the second-last layer of Alexnet. Each histogram corresponds to an example pair of complex and simplified images from Figure 1 of \cite{Tanaka2003}. Neurons were found that responded maximally to each of the complex stimuli. The blue histograms show the distribution of strength of responses to the corresponding simplified stimuli (normalized by strength of response to the most effective complex stimulus). As a control, the gray histograms show the distribution of strength of responses to other (non-matching) simplified stimuli. B) Fractions of neurons with equal or greater responses to simplified stimuli than complex stimuli. The blue lines correspond to the matching simplified stimulus for each complex stimulus, and the gray lines correspond to the other simplified stimuli. 
}
\label{fig:simplification}
\end{figure}

The image-simplification process was manual and not fully specified, so it is not clear how accurately it could be reproduced for CNN units. Instead, the CNNs were tested with twelve examples of complex and simplified images provided in \cite{Tanaka2003}. The CNN was first shown the twelve complex images and 56 additional images taken from ShapeNet (total 68 images, similar to the numbers of stimuli described in \cite{kobatake1994neuronal}). For each of the 12 complex images, units were identified that responded maximally or nearly maximally (within 5\% of the mean unit-wise maximum) to that image among the 68 stimuli. Figure \ref{fig:simplification}A shows distributions of response strengths of these units when shown the corresponding simplified images, normalized to the response response to the complex image. The responses to simplified images are broadly distributed, but in many cases, a substantial fraction of the units responded at least as strongly to the simplified image as to the complex image. An exception was that very few of the units that responded strongly to a person wearing a lab coat responded as strongly to its simplification (a black circle over a white circle). This exception was consistent across other Alexnet layers, and across the VGG-16 layers. 

There are infinite ways in which an image could potentially be simplified, so it is striking that the particular simplifications that elicited maximal responses from the IT neurons in \cite{Tanaka2003} also elicited maximal responses from some of the 4098 units in this layer of Alexnet. 

Figure \ref{fig:simplification}B shows the fractions of neurons in each layer of each network that responded maximally to the simplified images in \cite{Tanaka2003}. Interestingly, a few of the units in each layer responded maximally to simplifications of \emph{other} objects (e.g. a unit that responded maximally to a complex bread image may also have responded maximally to a simplified cat). However, the fractions of maximal responses to simplification of the most effective stimulus were uniformly greater. 

\section{Discussion}
Past studies \cite{Zeiler2014,Yamins2014,Khaligh-Razavi2014,Hong2016} have shown that activity in the next-to-last layer of object-classification CNNs is closely related to that in the primate inferotemporal cortex (IT). The current study examined properties of the tuning of individual neurons to a number of stimulus variations that have been used to understand representation in IT. The results highlight some specific similarities between the CNN units and IT neurons, including particularly the distributions of population response sparseness, and size-tuning bandwidth. A number of other tuning properties had both similarities and differences with IT. These include scale and translation invariance, orientation tuning, and responses to occlusion. Responses to clutter were also systematically different. The distributions of stimulus selectivity were quite different. Overall, in agreement with past studies, the similarities are striking, particularly in light of the many differences in computational mechanisms and details of learning between the visual cortex and the CNNs.

Others \cite{Yamins2014,Cadieu2014,Khaligh-Razavi2014} have found relationships between IT activity and CNNs. However, they did not report individual neuron tuning properties, so similarity with IT may have varied differently with respect to different tuning properties in these networks.

\subsection{Limitations}
This study has examined a small selection of tuning properties, and the responses of only two CNNs, although these CNNs are widely used and reasonably representative. A more comprehensive investigation would examine the effects of additional CNN structures and training procedures. Furthermore, although efforts were made to approximate the stimuli used in IT studies, the stimuli could not typically be reproduced exactly. Generally an exact analogy isn't possible. For example, the number of pixels per degree visual angle is not well defined in the networks. Also, the stimuli in the IT studies appeared and disappeared, and sometimes moved, and the IT neurons responded dynamically. In contrast, the CNNs receive single frames of input and produce a single corresponding response (without dynamics). CNNs responses are probably most closely related to the earliest (largely feedforward) IT responses. However, IT responses change constantly, and latencies vary across conditions, so it is not clear which time window of IT responses would be the most closely related. 

\subsection{Toward IT-Like CNNs}
The many similarities between CNNs and the ventral visual stream suggest an opportunity to further optimize CNNs to more closely approximate the ventral stream, perhaps in terms of size invariance and orientation tuning. \cite{Yamins2014} trained CNNs specifically to emulate IT neuron activity, but this approach did not yield better IT approximations than simply training the CNNs for object classification. However, this attempt may have been hampered by the small size of the neural dataset. Another approach would be to use an empirical model of cortical activity to produce a large (but approximate) set of labels, as in \cite{Tripp2016}. Further incorporation of physiological mechanisms such as \cite{Rubin2015} into CNNs may also be important. 

\section{Conclusion}
This work is a preliminary comparison of various stimulus tuning properties of CNN units to corresponding properties of IT neurons. The results show a full spectrum of close similarities, subtle differences, and large differences with respect to different tuning properties. This provides some missing detail on the specific relationships between CNN unit and IT neuron responses, and suggests how CNN unit activity could be modified in future work to more closely reflect the visual representation in IT. 

%
\IEEEpeerreviewmaketitle

\section*{Acknowledgment}

Victor Reyes Osorio extracted the stimulus images from \cite{Lehky2011}. Salman Khan provided comments on an earlier draft.



%
\bibliography{it-cnn}

\begin{thebibliography}{10}
\providecommand{\url}[1]{#1}
\csname url@samestyle\endcsname
\providecommand{\newblock}{\relax}
\providecommand{\bibinfo}[2]{#2}
\providecommand{\BIBentrySTDinterwordspacing}{\spaceskip=0pt\relax}
\providecommand{\BIBentryALTinterwordstretchfactor}{4}
\providecommand{\BIBentryALTinterwordspacing}{\spaceskip=\fontdimen2\font plus
\BIBentryALTinterwordstretchfactor\fontdimen3\font minus
  \fontdimen4\font\relax}
\providecommand{\BIBforeignlanguage}[2]{{%
\expandafter\ifx\csname l@#1\endcsname\relax
\typeout{** WARNING: IEEEtran.bst: No hyphenation pattern has been}%
\typeout{** loaded for the language `#1'. Using the pattern for}%
\typeout{** the default language instead.}%
\else
\language=\csname l@#1\endcsname
\fi
#2}}
\providecommand{\BIBdecl}{\relax}
\BIBdecl

\bibitem{Zeiler2014}
M.~D. Zeiler and R.~Fergus, ``{Visualizing and Understanding Convolutional
  Networks arXiv:1311.2901v3 [cs.CV] 28 Nov 2013},'' \emph{Computer
  Vision–ECCV 2014}, vol. 8689, pp. 818--833, 2014.

\bibitem{Krizhevsky2012}
A.~Krizhevsky, I.~Sutskever, and G.~E. Hinton, ``{ImageNet Classification with
  Deep Convolutional Neural Networks},'' in \emph{Advances in Neural
  Information Processing Systems}, 2012, pp. 1--9.

\bibitem{Yamins2014}
D.~L.~K. Yamins, H.~Hong, C.~F. Cadieu, E.~a. Solomon, D.~Seibert, and J.~J.
  Dicarlo, ``{Performance-optimized hierarchical models predict neural
  responses in higher visual cortex.}'' \emph{PNAS}, may 2014.

\bibitem{Serre2007}
T.~Serre, A.~Oliva, and T.~Poggio, ``{A feedforward architecture accounts for
  rapid categorization.}'' \emph{PNAS}, vol. 104, no.~15, pp. 6424--9, apr
  2007.

\bibitem{Rolls2012}
E.~T. Rolls, ``{Invariant Visual Object and Face Recognition: Neural and
  Computational Bases, and a Model, VisNet.}'' \emph{Frontiers in Computational
  Neuroscience}, vol.~6, no. June, p.~35, jan 2012.

\bibitem{Khaligh-Razavi2014}
S.~M. Khaligh-Razavi and N.~Kriegeskorte, ``{Deep Supervised, but Not
  Unsupervised, Models May Explain IT Cortical Representation},'' \emph{PLoS
  Computational Biology}, vol.~10, no.~11, 2014.

\bibitem{Markov2014}
N.~T. Markov, M.~M. Ercsey-Ravasz, a.~R. {Ribeiro Gomes}, C.~Lamy, L.~Magrou,
  J.~Vezoli, P.~Misery, A.~Falchier, R.~Quilodran, M.~a. Gariel, J.~Sallet,
  R.~Gamanut, C.~Huissoud, S.~Clavagnier, P.~Giroud, D.~Sappey-Marinier,
  P.~Barone, C.~Dehay, Z.~Toroczkai, K.~Knoblauch, D.~C. {Van Essen}, and
  H.~Kennedy, ``{A weighted and directed interareal connectivity matrix for
  macaque cerebral cortex.}'' \emph{Cerebral Cortex}, vol.~24, no.~1, pp.
  17--36, jan 2014.

\bibitem{Schwartz1983}
E.~L. Schwartz, R.~Desimone, T.~D. Albright, and C.~G. Gross, ``{Shape
  recognition and inferior temporal neurons},'' \emph{PNAS}, vol.~80, no.
  September, pp. 5776--5778, 1983.

\bibitem{Hong2016}
H.~Hong, D.~L.~K. Yamins, N.~J. Majaj, and J.~J. DiCarlo, ``{Explicit
  information for category-orthogonal object properties increases along the
  ventral stream},'' \emph{Nature Neuroscience}, vol.~19, no.~4, pp. 613--622,
  2016.

\bibitem{dicarlo2012does}
J.~J. DiCarlo, D.~Zoccolan, and N.~C. Rust, ``How does the brain solve visual
  object recognition?'' \emph{Neuron}, vol.~73, no.~3, pp. 415--434, 2012.

\bibitem{Robinson2015}
L.~Robinson and E.~T. Rolls, ``{Invariant visual object recognition:
  biologically plausible approaches},'' \emph{Biological Cybernetics}, vol.
  109, no. 4-5, pp. 505--535, 2015.

\bibitem{Khan2017}
S.~Khan and B.~Tripp, ``An empirical model of activity in macaque inferior
  temporal cortex,'' \emph{Neural Networks}, accepted.

\bibitem{Russakovsky2015}
O.~Russakovsky, J.~Deng, H.~Su, J.~Krause, S.~Satheesh, S.~Ma, Z.~Huang,
  A.~Karpathy, A.~Khosla, M.~Bernstein, A.~C. Berg, and L.~Fei-Fei, ``{ImageNet
  Large Scale Visual Recognition Challenge},'' \emph{International Journal of
  Computer Vision}, vol. 115, no.~3, pp. 211--252, 2015.

\bibitem{Cadieu2014}
C.~F. Cadieu, H.~Hong, D.~L.~K. Yamins, N.~Pinto, D.~Ardila, E.~A. Solomon,
  N.~J. Majaj, and J.~J. DiCarlo, ``{Deep Neural Networks Rival the
  Representation of Primate IT Cortex for Core Visual Object Recognition},''
  \emph{PLoS Computational Biology}, vol.~10, no.~12, 2014.

\bibitem{simonyan2014very}
K.~Simonyan and A.~Zisserman, ``Very deep convolutional networks for
  large-scale image recognition,'' \emph{arXiv preprint arXiv:1409.1556}, 2014.

\bibitem{Ding2015}
\BIBentryALTinterwordspacing
W.~Ding, R.~Wang, F.~Mao, and G.~Taylor, ``{Theano-based Large-Scale Visual
  Recognition with Multiple GPUs},'' in \emph{ICLR2015}, 2015. [Online].
  Available: \url{http://arxiv.org/abs/1412.2302}
\BIBentrySTDinterwordspacing

\bibitem{Lehky2011}
S.~R. Lehky, R.~Kiani, H.~Esteky, and K.~Tanaka, ``{Statistics of visual
  responses in primate inferotemporal cortex to object stimuli.}''
  \emph{Journal of Neurophysiology}, vol. 106, no.~3, pp. 1097--117, sep 2011.

\bibitem{Logothetis1995}
N.~K. Logothetis, J.~Pauls, and T.~Poggio, ``{Shape representation in the
  inferior temporal cortex of monkeys.}'' \emph{Current Biology}, vol.~5,
  no.~5, pp. 552--563, 1995.

\bibitem{Kovacs1995}
G.~Y. Kov{\'{a}}cs, R.~Vogels, and G.~a. Orban, ``{Selectivity of macaque
  inferior temporal neurons for partially occluded shapes.}'' \emph{The Journal
  of Neuroscience}, vol.~15, no. March 1995, pp. 1984--1997, 1995.

\bibitem{Tanaka2003}
K.~Tanaka, ``{Columns for complex visual object features in the inferotemporal
  cortex: clustering of cells with similar but slightly different stimulus
  selectivities.}'' \emph{Cerebral Cortex}, vol.~13, pp. 90--99, 2003.

\bibitem{hasselmo1989object}
M.~Hasselmo, E.~Rolls, G.~Baylis, and V.~Nalwa, ``Object-centered encoding by
  face-selective neurons in the cortex in the superior temporal sulcus of the
  monkey,'' \emph{Experimental Brain Research}, vol.~75, no.~2, pp. 417--429,
  1989.

\bibitem{Freiwald2010}
W.~Freiwald and D.~Y. Tsao, ``{Functional compartmentalization and viewpoint
  generalization within the macaque face-processing system},'' \emph{Science},
  vol. 330, pp. 845--851, 2010.

\bibitem{Ito1995}
M.~Ito, H.~Tamura, I.~Fujita, and K.~Tanaka, ``{Size and position invariance of
  neuronal responses in monkey inferotemporal cortex.}'' \emph{Journal of
  Neurophysiology}, vol.~73, no.~1, pp. 218--226, 1995.

\bibitem{OpDeBeeck2000}
H.~{Op De Beeck} and R.~Vogels, ``{Spatial sensitivity of macaque inferior
  temporal neurons},'' \emph{Journal of Comparative Neurology}, vol. 426, pp.
  505--518, 2000.

\bibitem{nielsen2006dissociation}
K.~J. Nielsen, N.~K. Logothetis, and G.~Rainer, ``Dissociation between local
  field potentials and spiking activity in macaque inferior temporal cortex
  reveals diagnosticity-based encoding of complex objects,'' \emph{The Journal
  of neuroscience}, vol.~26, no.~38, pp. 9639--9645, 2006.

\bibitem{Zoccolan2005}
D.~Zoccolan, D.~D. Cox, and J.~J. DiCarlo, ``{Multiple Object Response
  Normalization in Monkey Inferotemporal Cortex},'' \emph{Journal of
  Neuroscience}, vol.~25, no.~36, pp. 8150--8164, 2005.

\bibitem{kobatake1994neuronal}
E.~Kobatake and K.~Tanaka, ``Neuronal selectivities to complex object features
  in the ventral visual pathway of the macaque cerebral cortex,'' \emph{Journal
  of neurophysiology}, vol.~71, no.~3, pp. 856--867, 1994.

\bibitem{Tripp2016}
B.~Tripp, ``{A convolutional model of the primate middle temporal area},'' in
  \emph{ICANN}, 2016.

\bibitem{Rubin2015}
\BIBentryALTinterwordspacing
D.~B. Rubin, S.~D.~V. Hooser, and K.~D. Miller, ``{The stabilized supralinear
  network: A unifying circuit motif underlying multi-input integration in
  sensory cortex},'' \emph{Neuron}, vol.~85, no.~1, pp. 1--51, 2015. [Online].
  Available: \url{http://dx.doi.org/10.1016/j.neuron.2014.12.026}
\BIBentrySTDinterwordspacing

\end{thebibliography}
\bibliographystyle{IEEEtran}



\end{document}